\documentclass[12pt]{article}
\usepackage{amssymb}
\usepackage{graphicx}
\usepackage[cp1251]{inputenc}
\usepackage{rotating}

\begin{document}
 \noindent {\footnotesize\it Astrophysical Bulletin, 2026, Vol. 81, No. 2 }
 \newcommand{\dif}{\textrm{d}}

 \noindent
 \begin{tabular}{llllllllllllllllllllllllllllllllllllllllllllll}
 & & & & & & & & & & & & & & & & & & & & & & & & & & & & & & & & & & & & & &\\\hline\hline
 \end{tabular}

  \vskip 0.2cm
  \bigskip
\centerline{\Large\bf  Kinematic Properties of the TW Hya Association}

 \bigskip
  \centerline {V. V. Bobylev\footnote [1]{bob-v-vzz@rambler.ru}}
 \bigskip
 \centerline{\small\it Pulkovo Astronomical Observatory, Russian Academy of Sciences, St. Petersburg, 196140 Russia}
 \bigskip
{Abstract---A kinematic analysis of the young stellar association TWHya has been performed. The components of the displacement matrix in the Ogorodnikov-Milne linear model have been estimated both graphically and by solving the basic kinematic equations. The association's volume expansion with a coefficient of $K_{xyz}=103\pm9$~km s$^{-1}$ kpc$^{-1}$ was confirmed, which yields a dynamical age estimate of $t = 9.7 \pm0.8$~Myr. Using the graphical method, estimates of the association's proper rigid-body rotation parameters $\omega$ around the galactic axes $x $ and $y $ have been obtained for the first time, with velocity values
in the range of 50--70~km s$^{-1}$ kpc$^{-1}$ and errors in their determination of 14--19~km s$^{-1}$ kpc$^{-1}$. However,
these values are not confirmed by another method. For example, when solving kinematic equations only
using proper motions, all three components of rigid body rotation do not differ significantly from zero,
$(\omega_x,\omega_y,\omega_z)=(4,7,11)\pm(5,5,5)$~km s$^{-1}$ kpc$^{-1}$.
  }

 \bigskip
 \section{INTRODUCTION}
The young stellar association TW Hya is one of the closest to the Sun. Most of its members are lowmass stars of spectral types K and M that have not yet reached the main sequence stage.

A large number of publications have been devoted to the study of this association (de la Reza et al., 1989;
Gregorio-Hetem et al., 1992; Zuckerman and Becklin, 1993; Makarov and Fabricius, 2001; Mamajek,
2005; de la Reza et al., 2006; Ducourant et al., 2014; Donaldson et al., 2016; Luhman, 2023; Bobylev and
Bajkova, 2024; Olivares et al., 2025). The age of TW Hya is approximately 10 Myr (see, e.g., Gagn\'e et
al., 2018). It was estimated both on the basis of stellar kinematics (by the expansion effect) and by fitting to
suitable isochrones. It is worth noting the work of Miret-Roig et al. (2025), where Table C.1 provides an
extensive summary of historical age estimates for this association, demonstrating good agreement between
estimates obtained by various methods. All estimates fall within the range of 3--12 Myr.

The main interest in studying the kinematics of the TW Hya association is that it is quite compact, with its center at a distance of about 60 pc from the Sun. In this case, the velocity components of the association stars, calculated from the proper motions from the Gaia DR3 catalog (Vallenari et al., 2023), are very accurate. Indeed, the average error of the Gaia trigonometric parallaxes is 0.04 mas and 0.04 mas yr$^{-1}$ in the proper motion. Here, the average relative parallax error is 0.3\%, so the stars of the TW Hya association have an average error of rectangular coordinates of about 0.07 pc and about 0.06 km s$^{-1}$ for tangential velocities. Thus, it is desirable for the stars of this association to have radial velocity values with an average measurement error
significantly less than 1 km s$^{-1}$, which is not currently the case.

Kinematic analysis of stellar groups near the Sun using high-precision data currently yields important results. For example, Armstrong et al. (2025) studied the kinematic features of a large number of very young small stellar groups within the US OB association (part of the Scorpio-Centaurus OB association). The authors concluded that stochastic star formation occurs within the US. It was precisely for the US OB association that the sequential star formation model was developed (Preibisch and Zinnecker,
1999). A modification of this model is currently being applied to the US in the form of the ``cluster-to cluster''
model (Posch et al., 2025).

OB associations are known to be prone to expansion, which occurs after the most massive stars
explode as supernovae and sweep out a large amount of gas beyond the association. Quantitative confirmation
of this effect in the analysis of various OB associations was found in the papers of Mel'nik and
Dambis (2017, 2018), Wright (2020), Bobylev and Bajkova (2023, 2024), Armstrong et al. (2025). The
best known estimate is the flat expansion coefficient $K_{xy }=50$~km s$^{-1}$ kpc$^{-1}$, found by Blaauw (1964)
from an analysis of the kinematics of the Scorpio-Centaurus OB association. However, such estimates
are usually based on the study of either expansion along one direction or expansion in some plane. The
uniqueness of the TW Hya association is that it has been recorded to have a volumetric expansion
with a coefficient of $K_{xyz}=102\pm9$~km s$^{-1}$ kpc$^{-1}$ (Luhman, 2023).

In the work of Luhman (2023), ground-based measurements of radial velocities of candidate stars of
the TW Hya association were collected and analyzed. Nagananda et al. (2024) and Miret-Roig et al.
(2025) obtained new radial velocity measurements for a number of stars from the list of the most likely
members of the association compiled by Luhman (2023).

The key value of the volumetric expansion coefficient is that it allows one to obtain a dynamic
age estimate for an association independent of the isochrone method. In particular, the dynamic age
of the TW Hya association is approximately 10 Myr (e.g., Luhman, 2023; Bobylev and Bajkova, 2024; Olivares et al., 2025).

Furthermore, special attention should be paid to studying the association's proper rotation. Firstly,
such rotation may indicate that the parent cloud formed in a turbulent environment. Secondly, it is
necessary to ensure that various analytical methods yield consistent results. Therefore, searching for any proper rotation of the TW Hya association is one of the research areas in our work.

The aim of this paper is to study the three-dimensional kinematics of the TW Hya association using the latest radial velocity data for several members of the association. We use methods for estimating
the parameters of the Ogorodnikov-Milne linear model, allowing the sample to include both stars with
measured proper motions and radial velocities, as well as stars for which only proper motion data are available.

\section{METHODS}
In our studies, we use a rectangular coordinate system with the Sun at the center. The $x-$axis is
directed toward the galactic center, the $y$-axis toward the galactic rotation, and the $z$-axis toward
the north pole of the galaxy. $x=r\cos l\cos b,$ $y=r\sin l\cos b$ and $z=r\sin b,$ where $r=1/\pi$ is the
heliocentric distance to the star in kpc, which is calculated through its trigonometric parallax $\pi$ in mas.

From observations we know the radial velocity of the star $V_r $ and two projections of its tangential velocity:
$V_l=4.74r\mu_l\cos b$ and $V_b=4.74r\mu_b,$ directed along the galactic longitude l and latitude b, respectively,
expressed in km s$^{-1}$. Here the coefficient 4.74 is the ratio of the number of kilometers in an astronomical
unit to the number of seconds in a tropical year. The components of the proper motion $\mu_l\cos b$ and $\mu_b$ are expressed in mas yr$^{-1}$.

Using the components $V_r, V_l, V_b,$ the velocities $U,V,W$ are calculated, where the velocity $U$ is directed
from the Sun to the center of the Galaxy, $V$---in the direction of rotation of the Galaxy, and $W$---to the north galactic pole:
 \begin{equation}
 \begin{array}{lll}
 U=V_r\cos l\cos b-V_l\sin l-V_b\cos l\sin b,\\
 V=V_r\sin l\cos b+V_l\cos l-V_b\sin l\sin b,\\
 W=V_r\sin b                +V_b\cos b.
 \label{UVW}
 \end{array}
 \end{equation}
Thus, the velocities $U,V,W$ are directed along the corresponding coordinate axes $x, y, z$.

 \subsection{Ogorodnikov-Milne linear model}
In the linear Ogorodnikov-Milne model (Ogorodnikov, 1965), the observed velocity of a star ${\bf V}(r)$, which has a heliocentric radius vector  ${\bf r}$, is described with an accuracy of first-order terms $r/R_0\ll 1$ by an equation in vector form:
\begin{equation}
 {\bf V}(r)={\bf V}_\odot+M{\bf r}+{\bf V'},
 \label{eq-1}
 \end{equation}
where ${\bf V}_\odot(X_\odot,Y_\odot,Z_\odot)$ is the peculiar velocity of the Sun relative to the stars under consideration;
{$\bf V'$} is the residual velocity of the star; $M$ is the displacement matrix whose components are the partial derivatives
of the velocity ${\bf u}(u_1,u_2,u_3)$ with respect to the distance ${\bf r}(r_1, r_2, r_3)$. Here ${\bf u}={\bf V}(R)-{\bf V}(R_0)$, and $R$ and $R_0$ are the galactocentric distances of the star and the Sun (more precisely, the distances from the Sun
to the galactic rotation axis), respectively. Then the elements of the displacement matrix are defined as
\begin{equation}
 M_{pq}={\left(\frac{\partial u_p} {\partial r_q}\right)}_\circ, \quad p,q=1,2,3,
 \label{eq-2}
 \end{equation}
where the zero index means that the derivatives are taken at the point $R=R_0$. All nine elements of the
matrix $M$ are determined using three components of the observed velocities---the radial velocities $V_r$ and
the components $V_l, V_b$ calculated on the basis of the proper motions of the stars:
\begin{equation}
  \begin{array}{lll}
 V_r=-U_{\odot}\cos b\cos l-V_{\odot}\cos b\sin l -W_{\odot}\sin b\\
 +\cos^2 b\cos^2 l M_{11}  +r[\cos^2 b\cos l\sin l M_{12}+\cos b\sin b \cos l M_{13}\\
         +\cos^2 b\sin l\cos l M_{21}+\cos^2 b\sin^2 l M_{22}+\cos b\sin b\sin l M_{23}\\
 +\sin b\cos b\cos lM_{31}  +\cos b\sin b\sin l M_{32}+\sin^2 b M_{33}],\\
   \\
  V_l= U_\odot\sin l-V_\odot\cos l \\
    -\cos b\cos l\sin l  M_{11} +r[-\cos b\sin^2 l M_{12}  -\sin b \sin l M_{13}\\
    +\cos b\cos^2 l M_{21} +\cos b\sin l\cos l M_{22}+\sin b\cos l  M_{23}],\\
 \\
 V_b=U_\odot\cos l\sin b+ V_\odot\sin l\sin b -W_\odot\cos b\\
 -\sin b\cos b\cos^2 l M_{11} +r[-\sin b\cos b\sin l \cos l M_{12}-\sin^2 b \cos l  M_{13} \\
 -\sin b\cos b\sin l\cos l M_{21} -\sin b\cos b\sin^2 l  M_{22} -\sin^2 b\sin l  M_{23}\\
  +\cos^2 b\cos l M_{31}  +\cos^2 b\sin l M_{32}+ \sin b\cos b  M_{33}].
   \label{eq-5}
  \end{array}
 \end{equation}
The matrix $M$ can be divided into symmetric $M^{\scriptscriptstyle+}$  (local deformation tensor) and antisymmetric
$M^{\scriptscriptstyle-}$ (rotation tensor) parts:
 \begin{equation}
  \begin{array}{lll}
 \renewcommand{\arraystretch}{2.2}
 \displaystyle
 M_{\scriptstyle pq}^{\scriptscriptstyle+}= {1\over 2}\left( \frac{\partial u_{p}}{\partial r_{q}}+
 \frac{\partial u_{q}}{\partial r_{p}}\right)_\circ,  \qquad \quad
 M_{\scriptstyle pq}^{\scriptscriptstyle-}={1\over 2}\left(\frac{\partial u_{p}}{\partial r_{q}}-
 \frac{\partial u_{q}}{\partial r_{p}}\right)_\circ.
  \label{eq-6}
  \end{array}
\end{equation}
The quantities $M_{\scriptscriptstyle32}^{\scriptscriptstyle-},
 M_{\scriptscriptstyle13}^{\scriptscriptstyle-},  M_{\scriptscriptstyle21}^{\scriptscriptstyle-}$  are the components of the rigid-body rotation vector of the small circumsolar neighborhood around the $x, y, z$ axes, respectively. In accordance with the rectangular coordinate system we have chosen, positive rotations are considered to be rotations from axis 1 to 2 ($\omega_z$), from axis 2 to 3 ($\omega_x$), from axis 3 to 1 ($\omega_y$):
 \begin{equation}
 M^{\scriptscriptstyle-}= \pmatrix
  {         0&-\omega_z & \omega_y\cr
     \omega_z&         0&-\omega_x\cr
    -\omega_y& \omega_x&         0\cr}.
 \label{Omega-0}
 \end{equation}
Each of the quantities $M_{\scriptscriptstyle12}^{\scriptscriptstyle+},
  M_{\scriptscriptstyle13}^{\scriptscriptstyle+},
  M_{\scriptscriptstyle23}^{\scriptscriptstyle+}$ describes the deformation in the corresponding plane. The
diagonal components of the local deformation tensor
$M_{\scriptscriptstyle11}^{\scriptscriptstyle+},
  M_{\scriptscriptstyle22}^{\scriptscriptstyle+},
  M_{\scriptscriptstyle33}^{\scriptscriptstyle+}$ (as well as the off-diagonal components
of the matrix $M: M_{\scriptscriptstyle12},  M_{\scriptscriptstyle13},  M_{\scriptscriptstyle23}$) describe the
overall local compression or expansion of the entire stellar system. The system of conditional equations (4) includes twelve unknown variables, which can be found using the least-squares method (LSM).

  \begin{table}[p]
  \caption[]
   {\small  Coordinates, parallaxes and proper motions of stars in the TW Hya association. The columns of the Table
indicate: (1)---numbers of stars according to the Gaia~DR3 catalogue, (2) and (3)---equatorial coordinates of objects,
(4)---trigonometric parallaxes, (5) and (6)---proper motions of stars taken from the Gaia~DR3 catalogue }
  \begin{center}  \label{Table-1}    \small
  \begin{tabular}{|c|c|c|c|c|c|c|}\hline
 Gaia DR3 & $\alpha$, deg & $\delta$, deg & $\pi\pm\sigma$, mas & $\mu_\alpha\cos\delta\pm\sigma,$  mas& $\mu_\delta\pm\sigma,$  mas\\\hline
            (1)      &  (2) &  (3) &  (4) &  (5) & (6) \\\hline
3532218595001808768 & 167.8671 & $-26.9175$ & $20.27\pm0.08 $ & $ -83.82\pm0.07 $ & $-20.98\pm0.07$ \\
5396105586807802880 & 170.2725 & $-38.7547$ & $15.29\pm0.02 $ & $ -63.05\pm0.03 $ & $-14.60\pm0.02$ \\
5397574190745629312 & 171.7136 & $-38.4155$ & $14.62\pm0.02 $ & $ -60.70\pm0.03 $ & $-15.49\pm0.02$ \\
5378040370245563008 & 179.8658 & $-45.1721$ & $13.84\pm0.09 $ & $ -56.45\pm0.07 $ & $-18.34\pm0.07$ \\
3465989374664029184 & 180.6579 & $-33.4780$ & $15.98\pm0.03 $ & $ -66.24\pm0.03 $ & $-23.39\pm0.02$ \\
6145304323118631680 & 187.5214 & $-44.0434$ & $12.79\pm0.02 $ & $ -52.18\pm0.01 $ & $-21.90\pm0.01$ \\
6139584010795996160 & 192.7046 & $-42.5233$ & $10.26\pm0.04 $ & $ -38.72\pm0.04 $ & $-19.98\pm0.03$ \\
5412403269717562240 & 146.6154 & $-44.9613$ & $21.44\pm0.03 $ & $ -78.26\pm0.03 $ & $~~9.26\pm0.03$ \\
5460240959047928832 & 153.0376 & $-31.4126$ & $18.79\pm0.06 $ & $ -78.51\pm0.04 $ & $-11.59\pm0.06$ \\
5460240959050125568 & 153.0376 & $-31.4126$ & $18.79\pm0.06 $ & $ -78.51\pm0.04 $ & $-11.59\pm0.06$ \\
5414158429569765632 & 154.8231 & $-44.6267$ & $14.66\pm0.02 $ & $ -57.53\pm0.02 $ & $ -0.64\pm0.02$ \\
5416221633076680320 & 156.3369 & $-42.6983$ & $11.30\pm0.02 $ & $ -44.50\pm0.01 $ & $ -1.82\pm0.02$ \\
5467714064704570112 & 157.1905 & $-28.5105$ & $16.35\pm0.04 $ & $ -65.37\pm0.04 $ & $-12.56\pm0.04$ \\
5444751795151480320 & 160.6248 & $-33.6713$ & $29.33\pm0.03 $ & $-118.75\pm0.02 $ & $-19.65\pm0.03$ \\
5470330146463996032 & 162.3282 & $-25.1566$ & $ 9.50\pm0.04 $ & $ -39.56\pm0.04 $ & $ -8.38\pm0.04$ \\
3536988276442796800 & 164.7101 & $-23.7725$ & $22.83\pm0.03 $ & $ -95.38\pm0.03 $ & $-22.96\pm0.03$ \\
5401795662560500352 & 165.4659 & $-34.7048$ & $16.63\pm0.01 $ & $ -68.31\pm0.01 $ & $-13.90\pm0.01$ \\
5401822669314874240 & 165.5406 & $-34.5099$ & $16.88\pm0.13 $ & $ -69.49\pm0.12 $ & $-14.52\pm0.11$ \\
3532027383058513664 & 167.1829 & $-28.0808$ & $18.31\pm0.08 $ & $ -69.21\pm0.06 $ & $-22.18\pm0.07$ \\
5452498541764280832 & 167.3070 & $-30.0278$ & $21.77\pm0.20 $ & $ -85.80\pm0.20 $ & $-15.90\pm0.20$ \\
5396978667757576064 & 167.6157 & $-37.5311$ & $26.99\pm0.04 $ & $-115.52\pm0.04 $ & $-16.89\pm0.04$ \\
5396978667759696000 & 167.6157 & $-37.5311$ & $26.99\pm0.04 $ & $-115.52\pm0.04 $ & $-16.89\pm0.04$ \\
5399220743767211776 & 170.3214 & $-34.7794$ & $16.71\pm0.02 $ & $ -69.10\pm0.02 $ & $-17.96\pm0.02$ \\
5399220743767211264 & 170.3223 & $-34.7806$ & $16.71\pm0.02 $ & $ -69.07\pm0.02 $ & $-16.77\pm0.02$ \\
3534414590303807232 & 170.5216 & $-24.7776$ & $20.06\pm0.29 $ & $ -88.29\pm0.37 $ & $-41.11\pm0.19$ \\
5348165127505382400 & 171.0079 & $-52.8449$ & $12.38\pm0.03 $ & $ -47.85\pm0.03 $ & $ -7.82\pm0.03$ \\
5398663566250727680 & 172.9798 & $-34.6077$ & $20.13\pm0.06 $ & $ -84.79\pm0.06 $ & $-22.54\pm0.05$ \\
5398663566249861120 & 172.9798 & $-34.6077$ & $20.13\pm0.06 $ & $ -84.79\pm0.06 $ & $-22.54\pm0.05$ \\
3481965995873045888 & 173.0755 & $-30.3089$ & $21.42\pm0.22 $ & $ -89.41\pm0.25 $ & $-24.55\pm0.17$ \\
3481965141177021568 & 173.0758 & $-30.3312$ & $21.09\pm0.04 $ & $ -89.13\pm0.05 $ & $-25.22\pm0.03$ \\
3485098646237003136 & 173.1711 & $-26.8693$ & $21.63\pm0.03 $ & $ -91.04\pm0.03 $ & $-24.16\pm0.02$ \\
3485098646237003392 & 173.1715 & $-26.8657$ & $21.76\pm0.03 $ & $ -90.61\pm0.03 $ & $-27.38\pm0.02$ \\
3478940625208241920 & 174.8905 & $-30.6669$ & $20.45\pm0.03 $ & $ -86.42\pm0.02 $ & $-25.83\pm0.01$ \\
3478519134297202560 & 174.9626 & $-31.9894$ & $21.41\pm0.23 $ & $ -89.83\pm0.23 $ & $-25.34\pm0.16$ \\
3463395519358168064 & 177.0986 & $-37.4802$ & $13.14\pm0.03 $ & $ -57.04\pm0.03 $ & $-15.89\pm0.02$ \\
3463395523652894336 & 177.1006 & $-37.4804$ & $13.07\pm0.01 $ & $ -53.01\pm0.01 $ & $-18.33\pm0.01$ \\
3567379121431731328 & 180.0063 & $-17.5254$ & $18.85\pm0.03 $ & $ -78.94\pm0.03 $ & $-28.16\pm0.02$ \\
3465944500845668224 & 180.1143 & $-34.0938$ & $14.13\pm0.07 $ & $ -58.14\pm0.08 $ & $-21.02\pm0.05$ \\
3466327989885650176 & 181.7951 & $-32.5150$ & $12.24\pm0.06 $ & $ -46.67\pm0.06 $ & $-22.43\pm0.03$ \\
3466308095597260032 & 181.8637 & $-32.7835$ & $17.60\pm0.09 $ & $ -73.62\pm0.09 $ & $-26.57\pm0.05$ \\
3459372646830687104 & 181.8891 & $-39.5484$ & $15.46\pm0.12 $ & $ -64.04\pm0.09 $ & $-23.68\pm0.07$ \\
3459492631038236416 & 181.9511 & $-39.0014$ & $15.23\pm0.96 $ & $ -63.58\pm0.66 $ & $-24.47\pm0.69$ \\
6150861598484393856 & 183.8776 & $-39.8120$ & $18.66\pm0.03 $ & $ -76.85\pm0.02 $ & $-28.19\pm0.01$ \\
     \hline
     \end{tabular}
     \end{center}
     \end{table}
     \begin{table*}[t]
     \begin{center}
     Table 1. (Contd.)
  {\small
   \begin{tabular}{|c|c|c|c|c|c|c|}\hline
 Gaia DR3 & $\alpha$, deg & $\delta$, deg & $\pi\pm\sigma$, mas & $\mu_\alpha\cos\delta\pm\sigma,$  mas& $\mu_\delta\pm\sigma,$  mas\\\hline
            (1)      &  (2) &  (3) &  (4) &  (5) & (6) \\\hline
6151330196594603648 & 184.4964 & $-37.5788$ & $12.90\pm0.04 $ & $ -52.62\pm0.03 $ & $-21.80\pm0.03$ \\
6143632653128880896 & 184.5776 & $-45.4782$ & $12.32\pm0.03 $ & $ -50.41\pm0.03 $ & $-19.23\pm0.02$ \\
6132146982868270976 & 187.9083 & $-45.9833$ & $12.47\pm0.02 $ & $ -49.68\pm0.02 $ & $-20.95\pm0.01$ \\
6146107993101452160 & 188.7673 & $-41.6109$ & $17.40\pm0.03 $ & $ -69.52\pm0.02 $ & $-29.56\pm0.02$ \\
6147119548096085376 & 188.9536 & $-39.8403$ & $14.04\pm0.03 $ & $ -56.71\pm0.03 $ & $-24.86\pm0.03$ \\
6147117722735170176 & 189.0019 & $-39.8712$ & $14.15\pm0.02 $ & $ -59.06\pm0.02 $ & $-30.03\pm0.02$ \\
6147117727029871360 & 189.0040 & $-39.8696$ & $14.13\pm0.05 $ & $ -55.65\pm0.04 $ & $-23.88\pm0.04$ \\
6147044433411060224 & 189.3012 & $-40.3635$ & $15.72\pm0.03 $ & $ -62.89\pm0.02 $ & $-28.17\pm0.02$ \\
6132672029732817024 & 191.3087 & $-44.4856$ & $12.15\pm0.34 $ & $ -46.15\pm0.31 $ & $-22.52\pm0.24$ \\
6152893526035165312 & 191.9343 & $-38.2797$ & $11.94\pm0.35 $ & $ -44.73\pm0.34 $ & $-21.56\pm0.24$ \\
6183591791897683584 & 194.7623 & $-31.7550$ & $13.55\pm0.03 $ & $ -53.18\pm0.02 $ & $-28.65\pm0.02$ \\
3468438639892079360 & 186.7137 & $-33.2704$ & $15.54\pm0.10 $ & $ -63.04\pm0.12 $ & $-29.31\pm0.08$ \\
6132134304124539264 & 188.7342 & $-45.6355$ & $12.74\pm0.14 $ & $ -51.24\pm0.13 $ & $-26.76\pm0.12$ \\
6132134299824086144 & 188.7342 & $-45.6355$ & $12.74\pm0.14 $ & $ -51.24\pm0.13 $ & $-26.76\pm0.12$ \\
3534414594600352896 & 170.5216 & $-24.7776$ & $20.06\pm0.29 $ & $ -88.29\pm0.37 $ & $-41.11\pm0.19$ \\
6179256348830614784 & 196.5755 & $-34.4826$ & $11.82\pm0.10 $ & $ -45.35\pm0.09 $ & $-25.63\pm0.08$ \\
5457259083514583552 & 161.4690 & $-28.3251$ & $11.96\pm0.08 $ & $ -48.89\pm0.07 $ & $ -8.55\pm0.07$ \\
5401389770971149568 & 164.3194 & $-35.2153$ & $10.60\pm0.09 $ & $ -38.57\pm0.08 $ & $-10.76\pm0.08$ \\
5399990638128330752 & 166.6856 & $-37.2532$ & $~9.80\pm0.26 $ & $ -41.30\pm0.27 $ & $ -7.76\pm0.26$ \\
3493814268751183744 & 179.2006 & $-22.4894$ & $12.75\pm0.06 $ & $ -53.14\pm0.04 $ & $-19.88\pm0.03$ \\
3459725624422311424 & 180.9958 & $-38.3613$ & $12.17\pm0.18 $ & $ -50.24\pm0.13 $ & $-18.67\pm0.10$ \\
6145303429765430784 & 187.5239 & $-44.0756$ & $12.82\pm0.03 $ & $ -52.01\pm0.02 $ & $-21.73\pm0.02$ \\
6133420114251217664 & 189.2707 & $-44.3221$ & $11.14\pm0.06 $ & $ -44.41\pm0.04 $ & $-19.22\pm0.04$ \\
6114656192408518784 & 209.4920 & $-37.9930$ & $12.54\pm0.02 $ & $ -42.59\pm0.02 $ & $-31.55\pm0.03$ \\
     \hline
     \end{tabular} }
     \end{center}
     \end{table*}
  \begin{table}[p]
  \caption[]
   {\small  Radial velocities of stars in the TW Hya association. The columns of the Table contain: (1)---numbers of stars
according to the Gaia~DR3 catalogue, (2) and (3)---equatorial coordinates of objects, (4)---radial velocities of stars. In
column (5) the following designations of literary sources are introduced: [1]---Luhman (2023); [2]---Nagananda et al.
(2024); [3]---Miret-Roig et al. (2025) }
  \begin{center}  \label{Table-2}    \small
  \begin{tabular}{|c|c|c|c|c|c|c|}\hline
 Gaia DR3 & $\alpha$, deg & $\delta$, deg  &  $V_r\pm\sigma$,  km s$^{-1}$ ~\quad & Ref \\\hline
          1()        &      (2) &     (3) &   (4) &  (5) \\\hline
3532218595001808768 & 167.867062 & $-26.917533$ & $ 11.0\pm2.5 $ & 2 \\
5396105586807802880 & 170.272480 & $-38.754655$ & $ 11.6\pm1.0 $ & 2 \\
5397574190745629312 & 171.713550 & $-38.415451$ & $ 12.2\pm1.0 $ & 2 \\
5378040370245563008 & 179.865771 & $-45.172094$ & $ 11.6\pm2.0 $ & 2 \\
3465989374664029184 & 180.657946 & $-33.477954$ & $~7.5\pm2.0 $ & 2 \\
6145304323118631680 & 187.521409 & $-44.043422$ & $ 10.5\pm2.0 $ & 2 \\
6139584010795996160 & 192.704566 & $-42.523281$ & $ 11.0\pm2.0 $ & 2 \\

5412403269717562240 & 146.615442 & $-44.961309$ & $ 15.69\pm1.52 $ & 1 \\
5460240959047928832 & 153.037644 & $-31.412610$ & $ 15.14\pm1.66 $ & 1 \\
5460240959050125568 & 153.037644 & $-31.412610$ & $ 14.08\pm0.81 $ & 1 \\
5414158429569765632 & 154.823077 & $-44.626664$ & $ 15.84\pm1.06 $ & 1 \\
5416221633076680320 & 156.336889 & $-42.698328$ & $ 17.87\pm 0.11 $ & 1 \\
5467714064704570112 & 157.190500 & $-28.510480$ & $ 12.40\pm 0.30 $ & 1 \\
5444751795151480320 & 160.624790 & $-33.671262$ & $ 12.45\pm 0.01 $ & 1 \\
5470330146463996032 & 162.328173 & $-25.156608$ & $ 19.0 \pm2.0  $ & 1 \\
3536988276442796800 & 164.710121 & $-23.772498$ & $ ~8.20\pm 0.20 $ & 1 \\
5401795662560500352 & 165.465903 & $-34.704793$ & $ 12.50\pm0 .02 $ & 1 \\
5401822669314874240 & 165.540616 & $-34.509943$ & $ ~9.0 \pm3.0 ~ $ & 1 \\
3532027383058513664 & 167.182943 & $-28.080752$ & $~9.30\pm 0.48 $ & 1 \\
5452498541764280832 & 167.307030 & $-30.027840$ & $ 11.10\pm 0.01 $ & 1 \\
5396978667757576064 & 167.615653 & $-37.531057$ & $~9.89\pm 0.62 $ & 1 \\
5396978667759696000 & 167.615653 & $-37.531057$ & $~9.52\pm0.86 $ & 1 \\
5399220743767211776 & 170.321373 & $-34.779386$ & $ 11.67\pm0.07 $ & 1 \\
5399220743767211264 & 170.322306 & $-34.780568$ & $ 12.07\pm0.04 $ & 1 \\
3534414590303807232 & 170.521605 & $-24.777618$ & $~8.70\pm 0.90 $ & 1 \\
5348165127505382400 & 171.007879 & $-52.844865$ & $~7.67\pm4.06 $ & 1 \\
5398663566250727680 & 172.979796 & $-34.607657$ & $ 13.4 \pm2.0 ~ $ & 1 \\
5398663566249861120 & 172.979796 & $-34.607657$ & $ 13.43\pm1.35 $ & 1 \\
3481965995873045888 & 173.075452 & $-30.308899$ & $ 12.0 \pm3.0~  $ & 1 \\
3481965141177021568 & 173.075839 & $-30.331184$ & $ 12.30\pm1.50 $ & 1 \\
3485098646237003136 & 173.171102 & $-26.869304$ & $ ~8.61\pm 0.03 $ & 1 \\
3485098646237003392 & 173.171490 & $-26.865667$ & $~8.68\pm 0.02 $ & 1 \\
3478940625208241920 & 174.890491 & $-30.666883$ & $ ~5.80\pm 0.70 $ & 1 \\
3478519134297202560 & 174.962604 & $-31.989403$ & $ 11.6 \pm2.0~ $ & 1 \\
3463395519358168064 & 177.098576 & $-37.480215$ & $ 12.28\pm 0.03 $ & 1 \\
3463395523652894336 & 177.100631 & $-37.480391$ & $ 11.65\pm 0.02 $ & 1 \\
3567379121431731328 & 180.006251 & $-17.525365$ & $ 12.29\pm4.15 $ & 1 \\
3465944500845668224 & 180.114282 & $-34.093756$ & $ 11.0 \pm2.0~ $ & 1 \\
3466327989885650176 & 181.795122 & $-32.515032$ & $ 10.47\pm0 .41 $ & 1 \\
3466308095597260032 & 181.863679 & $-32.783519$ & $  ~7.71\pm2.61 $ & 1 \\
3459372646830687104 & 181.889079 & $-39.548443$ & $ 11.2 \pm2.0~  $ & 1 \\
3459492631038236416 & 181.951108 & $-39.001354$ & $  20  \pm7~ $ & 1 \\
6150861598484393856 & 183.877557 & $-39.811960$ & $ ~6.14\pm1.68 $ & 1 \\
 \hline
 \end{tabular}\end{center}
 \end{table}
     \begin{table*}[t]
     \begin{center}
     Table 2. (Contd.)

  {\small
   \begin{tabular}{|c|c|c|r|c|c|c|}\hline
 Gaia DR3 & $\alpha$, deg & $\delta$, deg  &  $V_r\pm\sigma$,  km s$^{-1}$ ~\quad & Ref \\\hline
          1()        &      (2) &     (3) &   (4) &  (5) \\\hline
6151330196594603648 & 184.496373 & $-37.578796$ & $ 13.72\pm2.40 $ & 1 \\
6143632653128880896 & 184.577591 & $-45.478248$ & $ ~9.34\pm3.01 $ & 1 \\
6132146982868270976 & 187.908304 & $-45.983276$ & $  ~8.1\pm4.0~$ & 1 \\
6146107993101452160 & 188.767320 & $-41.610858$ & $  ~6.31\pm0 .23 $ & 1 \\
6147119548096085376 & 188.953587 & $-39.840268$ & $  ~8.43\pm1.47 $ & 1 \\
6147117722735170176 & 189.001946 & $-39.871159$ & $  ~8.92\pm 0.06 $ & 1 \\
6147117727029871360 & 189.003977 & $-39.869612$ & $ 10.95\pm 0.59 $ & 1 \\
6147044433411060224 & 189.301235 & $-40.363514$ & $  ~6.30\pm 0.90 $ & 1 \\
6132672029732817024 & 191.308716 & $-44.485580$ & $  ~8.0 \pm3.0~  $ & 1 \\
6152893526035165312 & 191.934281 & $-38.279658$ & $     14\pm6~~~ $ & 1 \\
6183591791897683584 & 194.762347 & $-31.755036$ & $  ~3.79\pm3.82 $ & 1 \\
3468438639892079360 & 186.713655 & $-33.270355$ & $  ~7.15\pm 0.26 $ & 1 \\
6132134304124539264 & 188.734159 & $-45.635501$ & $  ~7.07\pm3.21 $ & 1 \\
6132134299824086144 & 188.734159 & $-45.635501$ & $  ~9.27\pm 0.38 $ & 1 \\
3534414594600352896 & 170.521605 & $-24.777618$ & $  ~5.70\pm 0.10 $ & 3 \\
 \hline
 \end{tabular} }
 \end{center}
  \end{table*}

 \subsection{Graphical method}
All nine elements of the displacement matrix $M$ can also be determined graphically. To do this, it is necessary to find a linear relationship of the form $v=M_{pq}\cdot x+b, ~ p,q=1,2,3$, for example, using the least-squares method. To apply
this method, the spatial velocities of the stars are required: $U,V,W$. According to relations (1), their calculation requires the combined use of trigonometric parallaxes, proper motions, and radial velocities of the stars.

It was using a graphical method that Luhman (2023) estimated the coefficient of volumetric expansion
of the TWA Hya association:
$K_{xyz}=(\partial U/\partial x+\partial V/\partial y+\partial W/\partial z)/3=102\pm9$~km s$^{-1}$ kpc$^{-1}$.

Moreover, the values of all three gradients turned out to be significantly different from zero. In the
work of Bobylev and Bajkova (2024), the obtained value of $K_{xyz}$ was confirmed using the same graphical
method---using practically the same data, the authors
found $K_{xyz}= 103\pm12$~km s$^{-1}$ kpc$^{-1}$.

Estimating the values of the elements of the displacement matrix $M$ by solving a system of kinematic
equations of the form (4) is of interest for comparing results obtained by different methods. In this case,
all nine elements of the matrix $M $ are determined simultaneously as a result of a joint solution.

\section{DATA}
In this paper, we use the list of members of the TW Hya association compiled by Luhman (2023).
This list contains 67 of the most likely members of
the association, which are members of 55 binary or
multiple systems. Not all stars have radial velocity
measurements. We expanded the list of stars in the
association with measured radial velocities to 58 by adding data from Nagananda et al. (2024) and Miret-Roig et al. (2025).

 Nagananda et al. (2024) refined the lists of candidate stars in the following seven young stellar associations:
TW Hya, Tuc-Hor, Argus, $\beta$ Pic, Carina, Columba, and AB Dor. For seven stars from the TW Hya association list (Luhman, 2023), for which radial velocity measurements were not previously available, the authors used spectroscopic data,
which allowed them to estimate the radial velocities with an acceptable accuracy (less than 2 km s$^{-1}$).

Olivares et al. (2025) conducted a detailed study of the distribution of stars in the TW Hya association and discovered an age gradient. The authors found that the association consists of two parts with different ages: one part has an age of 9 Myr, the
other---6 Myr. In this case, the ages were estimated
from photometric data using an isochronous fitting.
Olivares et al. (2025) confirmed a small difference
in the ages of these groupings using a dynamical
method. In addition, the authors showed an evolutionary relationship of the TW Hya association with
an older $\sigma$ Cen grouping, as well as with the even older Scorpio-Centaurus OB association. For the analyzed stars, the authors collected radial velocity data from the literature; we used them for the star Gaia~DR3 3534414594600352896.

The initial data on the stars in our sample are presented in Tables 1 and 2.

 \section{RESULTS AND DISCUSSION}
\subsection{Results of applying the graphical method}
Figure 1 shows the velocities $U, V,W$ for 58 stars in the TW Hya association as functions of the corresponding
$x, y, z$ coordinates. Using the data from each of the nine graphs, a search wasmade for a linear dependence of the form
$v=M_{pq}\cdot x+b, ~~ p,q=1,2,3$
using the least-squares method. In each case, large residuals were rejected using the 3$\sigma$ criterion. As a
result, the following parameters were found:
\begin{equation}
 \begin{array}{lll}
  M_{11}=\partial U/\partial x = 112\pm8~\hbox{km s$^{-1}$ kpc$^{-1}$},   ~\quad b=-14.0\pm0.2~\hbox{km s$^{-1}$},\\
  M_{12}=\partial U/\partial y =-60\pm17~\hbox{km s$^{-1}$ kpc$^{-1}$}, ~\quad b=-15.0\pm1.0~\hbox{km s$^{-1}$},\\
  M_{13}=\partial U/\partial z =165\pm31~\hbox{km s$^{-1}$ kpc$^{-1}$},  ~\quad b=-16.8\pm0.8~\hbox{km s$^{-1}$}, \\\\

  M_{21}=\partial V/\partial x = -49\pm20~\hbox{km s$^{-1}$ kpc$^{-1}$},  ~\quad  b=-17.8\pm0.5~\hbox{km s$^{-1}$},\\
  M_{22}=\partial V/\partial y =  ~~87\pm17~\hbox{km s$^{-1}$ kpc$^{-1}$},  ~\quad  b=-14.2\pm0.9~\hbox{km s$^{-1}$},\\
  M_{23}=\partial V/\partial z =-145\pm37~\hbox{km s$^{-1}$ kpc$^{-1}$}, ~\quad  b=-15.1\pm0.8~\hbox{km s$^{-1}$}, \\\\

  M_{31}=\partial W/\partial x = ~~30\pm16~\hbox{km s$^{-1}$ kpc$^{-1}$}, ~\quad  b=-6.7\pm0.4~\hbox{km s$^{-1}$},\\
  M_{32}=\partial W/\partial y =-24\pm17~\hbox{km s$^{-1}$ kpc$^{-1}$}, ~\quad  b=-7.5\pm0.9~\hbox{km s$^{-1}$},\\
  M_{33}=\partial W/\partial z =123\pm25~\hbox{km s$^{-1}$ kpc$^{-1}$}, ~\quad  b=-9.1\pm0.6~\hbox{km s$^{-1}$}.
  \label{rez-1} \end{array}
 \end{equation}
In six cases, $M_{pq}$ differ significantly from zero. These dependencies are shown in the corresponding
panels of Fig. 1. The most interesting are the three gradients, which in the Ogorodnikov-Milne linear
model (Ogorodnikov 1965; Bobylev, Bajkova 2023) are the diagonal terms of the deformation matrix and
describe the effects of the expansion of the stellar system. In Fig. 1, the corresponding dependencies are indicated by red lines. Let us estimate the value of the volume expansion coefficient of the TW Hya association:
\begin{equation}
K_{xyz}=107\pm10~\hbox {km s$^{-1}$ kpc$^{-1}$}
 \label{t-xyz}
 \end{equation}
We will also find the time interval that has passed from the beginning of the expansion of the star system to the presentmoment: $t=977.5/K_{xyz}=9.1\pm1.1$~Myr.

Based on the found values (6) using relations (5),
we obtain the following values of the three angular velocities of rotation:
\begin{equation}
 \begin{array}{lll}
  \omega_x=M_{\scriptscriptstyle32}^{\scriptscriptstyle-}=61\pm18~\hbox{km s$^{-1}$ kpc$^{-1}$},\\
  \omega_y=M_{\scriptscriptstyle13}^{\scriptscriptstyle-}=68\pm17~\hbox{km s$^{-1}$ kpc$^{-1}$},\\
  \omega_z=M_{\scriptscriptstyle21}^{\scriptscriptstyle-}=~6\pm13~\hbox{km s$^{-1}$ kpc$^{-1}$}.
 \label{rez-rotat-1} \end{array}
 \end{equation}

\begin{figure}[t]
{ \begin{center}  \includegraphics[width=0.95\textwidth]{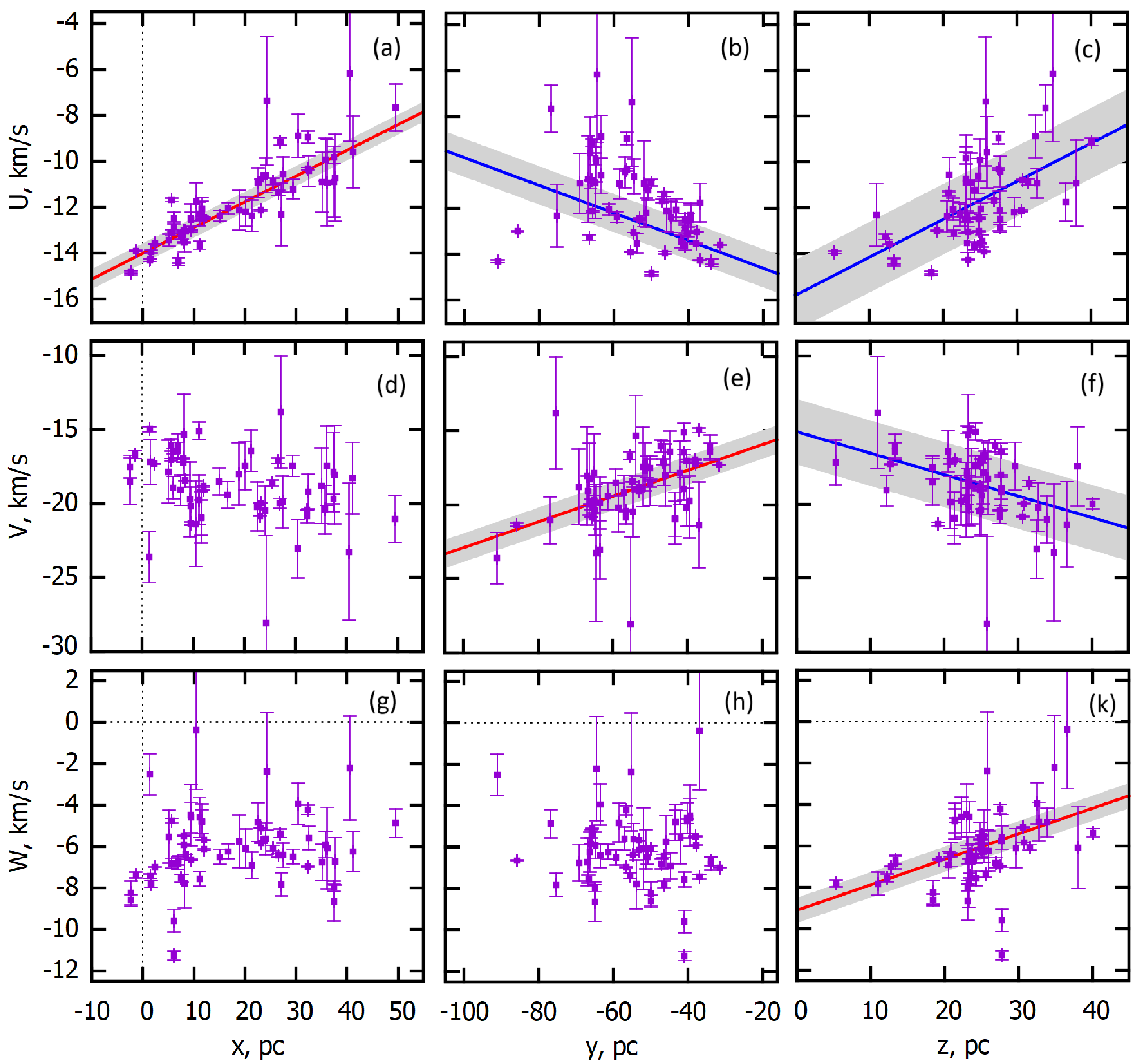}
  \caption{Dependences of velocities $U,V,W$ on coordinates $x,y,z$.} \label{f1}
\end{center}}\end{figure}

 \subsection{Application of the Ogorodnikov-Milne model}
The systemof conditional equations of the form~(4) was solved by the least squares method for three
cases. In the first case, 58 stars were used for which complete information was available---the radial velocity
and two components of proper motion. In the second case, all 67 stars from our list were used, all
58 stars with measured radial velocities, and each star without measured radial velocities gave only two
equations for $V_l$ and $V_b.$ Finally, in the third case, all 67 stars were used, but for stars with radial velocity
errors greater than 3 km s$^{-1}$, only their $V_l $ and $V_b$ components were used. The results are reflected in
Table~3, where Nequat is the number of equations used in the solved system of the form (4), and $\sigma_0$ is the unit
weight error obtained in the LSM-solution.

 Note that in both the first and second cases, the star Gaia~DR3 3459492631038236416 was rejected
by the 3$\sigma$ criterion. In the third case, the contribution of its radial velocity was not automatically taken into
account, since the measurement error here is very large---7 km s$^{-1}$.

Based on the data from the second column of Table 3, we obtain the following estimates of the three
angular velocities of rotation:
\begin{equation}
 \begin{array}{lll}
  \omega_x=58\pm14~\hbox{km s$^{-1}$ kpc$^{-1}$},\\
  \omega_y=16\pm13~\hbox{km s$^{-1}$ kpc$^{-1}$},\\
  \omega_z=-7\pm10~\hbox{km s$^{-1}$ kpc$^{-1}$}.
 \label{rez-rotat-5} \end{array}
 \end{equation}

{\begin{table}[t]                                                
\caption[]{\small\baselineskip=1.0ex\protect
Values of the parameters of the Ogorodnikov-Milne model}
\begin{center}\small
\label{t-Ogor}
\begin{tabular}{|l|r|r|r|r|r|}\hline
       Parameters &    58 stars & 67 stars & 67 stars \\\hline
$N_{equat}$        &            173 & 191 & 181\\
$\sigma_0$, km s$^{-1}$ &          1.32 & 1.27 & 1.04\\
           &&&  \\
 $U_\odot$, km s$^{-1}$ & $14.63\pm0.87$ &  $ 14.61\pm0.72$ &  $ 14.67\pm0.59$\\
 $V_\odot$, km s$^{-1}$ & $13.27\pm0.88$ & $ 13.36\pm0.82$ &  $ 13.06\pm0.67$\\
 $W_\odot$, km s$^{-1}$ & $ 8.48\pm0.87$ & $ 8.68\pm0.74$ &  $  8.87\pm0.62$\\
           &&&  \\
 $M_{11}$, km s$^{-1}$ kpc$^{-1}$  & $ 116\pm16$ &  $ 115\pm13$ &  $ 125\pm11$\\
 $M_{12}$, km s$^{-1}$ kpc$^{-1}$  & $    9\pm16$ &  $   7\pm13$ &  $ -5\pm11$\\
 $M_{13}$, km s$^{-1}$ kpc$^{-1}$  & $  44\pm26$ &  $ 41\pm22$ &  $ 13\pm19$\\

 $M_{21}$, km s$^{-1}$ kpc$^{-1}$   & $   -6\pm16$  & $ -6\pm15$ &  $ -22\pm13$\\
 $M_{22}$, km s$^{-1}$ kpc$^{-1}$   & $ 49\pm16$  &  $ 50\pm15$ &  $ 89\pm14$\\
 $M_{23}$, km s$^{-1}$ kpc$^{-1}$   & $-108\pm26$  &  $ -102\pm24$ &  $ -26\pm23$\\

 $M_{31}$, km s$^{-1}$ kpc$^{-1}$   & $  11\pm16$  & $ 10\pm13$ &  $ 17\pm11$\\
 $M_{32}$, km s$^{-1}$ kpc$^{-1}$   & $ 16\pm17$  &  $ 15\pm14$ &  $ -1\pm11$\\
 $M_{33}$, km s$^{-1}$ kpc$^{-1}$   & $119\pm26$  &  $ 127\pm23$ &  $ 95\pm20$\\
          &&&  \\
$K_{xyz}$, km s$^{-1}$ kpc$^{-1}$   & $95\pm12$ & $97\pm10$ &  $ 103\pm9$\\\hline
\end{tabular}
\end{center}
\end{table}}

A comparison of the kinematic analysis results for the TW Hya association stars, obtained by two methods, confirms the presence of a volumetric expansion for this association. Furthermore, there is a significantly non-zero proper rotation of the association around the galactic $x$-axis with an angular velocity of $\omega_x=58\pm14$~km s$^{-1}$ kpc$^{-1}$, which is consistent with the result (8) obtained graphically.

Our estimates of the group velocity components $U_\odot,V_\odot,W_\odot$  are close to the currently accepted
values of the peculiar velocity of the Sun relative
to the local standard of rest: $(U_\odot,V_\odot,W_\odot)=(11.1,12.2,7.3)$~km s$^{-1}$ (Sch\"onrich et al., 2010).
This coincidence seems surprising for such a close stellar grouping. It indicates that the average motion
of the association is not subject to any significant perturbations.

All parameters obtained by excluding radial velocities with large measurement errors (see the fourth column of Table 3) are determined with the smallest (compared to the other solutions) errors. Here, the
value of the expansion coefficient is $K_{xyz}=103\pm9$~km s$^{-1}$ kpc$^{-1}$. In this case, the
dynamical estimate of the age of the association is $t=9.7\pm0.8$~Myr.

The values of $K_{xyz}$ and $t $ found in the present work are in good agreement with the data presented in
the studies of Luhman (2023), Bobylev and Bajkova (2024). They also correspond to the estimates of
Olivares et al. (2025), although the authors of this work applied the dynamical method to two groups
of the TW Hya association (A and B) and obtained $t=10.2\pm1.0$~Myr and $t=8.5\pm1.3$~Myr for them,
respectively.

 Note that all three diagonal terms of the displacement matrix are significantly different from zero.
Therefore, in each of the three planes, there is a flat expansion effect with the following coefficients:
$K_{xy}=107\pm9$~km s$^{-1}$ kpc$^{-1}$,
$K_{yz}= 92\pm12$~km s$^{-1}$ kpc$^{-1}$ and
$K_{xz}=110\pm11$~km s$^{-1}$ kpc$^{-1}$.

 As can be seen from the last column of Table 3, the value of the unit weight error $\sigma_0$ (which is the
average of the sum of squared residuals) is close to 1~km s$^{-1}$. In fact, this value characterizes the internal
velocity dispersion in the TW Hya association. It can be noted that this value agrees well with the estimate
$0.8^{+0.3}_{-0.2}$~km s$^{-1}$ obtained in Mamajek (2005).

 Using the parameter values from column (4) of Table 3, we find:
 \begin{equation}
 \begin{array}{lll}
  \omega_x=13\pm13~\hbox{km s$^{-1}$ kpc$^{-1}$},\\
  \omega_y=-2\pm11~\hbox{km s$^{-1}$ kpc$^{-1}$},\\
  \omega_z=-9\pm9~\hbox{km s$^{-1}$ kpc$^{-1}$},
 \label{rez-rotat-55} \end{array}
 \end{equation}
which indicates the absence of any significant proper rotation of the TW Hya association around any axis.

It was decided to reapply the graphical method to a sample of stars with radial velocity errors less than
3~km s$^{-1}$. There were 47 such stars. As before, the 3$\sigma$ criterion was used to reject large residuals.
However, in this case, no errors were rejected by this criterion. Ultimately, the following parameters
were found (excluding the constant term, the values of which are of no interest here):
\begin{equation}
 \begin{array}{lll}
  M_{11}= 128\pm8~\hbox{km s$^{-1}$ kpc$^{-1}$}, \\
  M_{12}=-55\pm17~\hbox{km s$^{-1}$ kpc$^{-1}$},  \\
  M_{13}=173\pm31~\hbox{km s$^{-1}$ kpc$^{-1}$},\\ \\

  M_{21}= -84\pm19~\hbox{km s$^{-1}$ kpc$^{-1}$}, \\
  M_{22}= 106\pm16~\hbox{km s$^{-1}$ kpc$^{-1}$}, \\
  M_{23}=-128\pm34~\hbox{km s$^{-1}$ kpc$^{-1}$},\\ \\

  M_{31}= 40\pm15~\hbox{km s$^{-1}$ kpc$^{-1}$},\\
  M_{32}=-28\pm17~\hbox{km s$^{-1}$ kpc$^{-1}$},\\
  M_{33}=98\pm22~\hbox{km s$^{-1}$ kpc$^{-1}$}.\\
 \label{rez-777} \end{array}
 \end{equation}
 From these data we find estimates of three angular
rotation velocities:
\begin{equation}
 \begin{array}{lll}
  \omega_x= 50\pm19~\hbox{km s$^{-1}$ kpc$^{-1}$},\\
  \omega_y= 66\pm17~\hbox{km s$^{-1}$ kpc$^{-1}$},\\
  \omega_z=-15\pm13~\hbox{km s$^{-1}$ kpc$^{-1}$},
 \label{rez-rotat-111} \end{array}
 \end{equation}
 the values of which are close to those obtained by this method in solution (8).

The values of $\omega_z= 6\pm13$~km s$^{-1}$ kpc$^{-1}$, found in solution (8),
                       $\omega_z=-7\pm10$~km s$^{-1}$ kpc$^{-1}$ in (9),
                       $\omega_z=-9\pm9$~km s$^{-1}$ kpc$^{-1}$ in (10) and
                       $\omega_z=-15\pm13$~km s$^{-1}$ kpc$^{-1}$ in (12), are consistent with each other and indicate the absence of a rotation around the $z$-axis that is significantly different from zero.

Note that in the circumsolar neighborhood under consideration, the random errors of each component,
$V_l$ and $V_b$, are significantly smaller than the errors in the radial velocities $V_r$. Thus, the average errors of
$V_l$ and $V_b$ calculated using 67 stars are 0.1 and 0.07~km s$^{-1}$, respectively, and the average error of $V_r$ is
1.35~km s$^{-1}$. Even when using radial velocities with errors less than 3~km s$^{-1}$, the average error of $V_r$ is
0.93~km s$^{-1}$. Therefore, equations of the following form were used to estimate the three components of
the rigid body rotation vector:
  \begin{equation}
  \begin{array}{lll}
  V_l= U_\odot\sin l-V_\odot\cos l  +r[ -\cos l\sin b  ~\omega_{x} -\sin l\sin b ~\omega_{y}
     +\cos b ~\omega_{z}],\\
 V_b=U_\odot\cos l\sin b+ V_\odot\sin l\sin b - W_\odot\cos b+r[ \sin l  ~\omega_{x}+\cos l ~\omega_{y}],
   \label{eq-555}
  \end{array}
 \end{equation}
containing only six unknowns---three linear velocities of group motion $U_\odot, V_\odot, W_\odot$ and three angular velocities
of rigid body rotation $\omega_x, \omega_y, \omega_z$.

As a result of the LSM-solution of the system of conditional equations of form (13), the components of the group velocity of this stellar system were found, $(U,V,W)_\odot=(14.66,12.98,8.45)\pm(0.28,0.38,0.34)$~km s$^{-1}$, as well as the following components of the rotation vector:
 \begin{equation}
 \begin{array}{lll}
  \omega_x=  5\pm5~\hbox{km s$^{-1}$ kpc$^{-1}$},\\
  \omega_y=  7\pm5~\hbox{km s$^{-1}$ kpc$^{-1}$},\\
  \omega_z= 11\pm5~\hbox{km s$^{-1}$ kpc$^{-1}$}.
 \label{rez-rotat-777} \end{array}
 \end{equation}
All three components of the rotation vector (14) were determined with significantly smaller errors compared
to solution (10). The obtained data confirm the conclusion
previously drawn based on an analysis of the results of solving the main kinematic equations, regarding
the absence of significant proper rotation of the TW Hya association around any of the axes.

\section{CONCLUSIONS}
A comprehensive three-dimensional kinematic analysis of the young stellar association TW Hya was
carried out. The sample was formed from the most probable members of this association according to
the list from Luhman (2023). It consists of 67 stars with trigonometric parallaxes, proper motions from
the Gaia~DR3 catalog, and radial velocities collected from literature sources. Moreover, radial velocities
are known only for 58 stars. Luhman (2023) collected radial velocity measurements for 53 stars. In the
present study, the list of stars in the association with measured radial velocities was expanded to 58 by
adding measurements from Nagananda et al. (2024) and Miret-Roig et al. (2025).

 The objective of studying the kinematics of stars in the TW Hya association was to estimate the values
of nine terms of the displacement matrix in the Ogorodnikov-Milne linear model. This problem was
solved in two ways: graphically and by jointly solving the fundamental kinematic equations. Several discrepancies were found, but in most cases, good agreement was observed in the estimates of the model parameters.

The results obtained by both methods led to the conclusion that the volume expansion coefficient of
the TW Hya association can be reliably determined using either method. It was found with the smallest
error by solving a system of kinematic equations using stellar radial velocities measured with errors
of less than 3~km s$^{-1}$. In this case, the value of the coefficient $K_{xyz}$ is $103\pm9$~km s$^{-1}$ kpc$^{-1}$. The dynamic age estimate obtained for the association is $9.7\pm0.8$~Myr. Each plane exhibits a flat expansion effect with values indicating nearly isotropic expansion of the association.

Thus, the use of a number of additional measurements of stellar radial velocities in the present work
made it possible to obtain new estimates: by the graphical method---$K_{xyz}=107\pm10$~km s$^{-1}$ kpc$^{-1}$
($t=9.1\pm1.1$~Myr); even more accurately---from the solution of kinematic equations---$K_{xyz}=103\pm9$~km s$^{-1}$ kpc$^{-1}$ ($t=9.7\pm0.8$~Myr). The obtained values have smaller errors compared to the results
of Bobylev and Bajkova (2024): $K_{xyz}=103\pm12 $~km s$^{-1}$ kpc$^{-1}$ ($t=9.5\pm1.1$~Myr).
However, the main interest in this work is related to the estimation of the association rotation parameters.

There is complete agreement between the methods regarding the absence of proper rotation of the
TW Hya association around the galactic $z$-axis. Using the graphical method, we were the first to discover
significantly nonzero parameters describing the effects of rigid-body rotation $\omega$ of the TW Hya association
around the galactic $x$- and $y$-axes. The specific values of these velocities are 50--70~km s$^{-1}$ kpc$^{-1}$,
and the errors in their determination are 14--19~km s$^{-1}$ kpc$^{-1}$. However, these values are not
confirmed by another method. For example, when
solving kinematic equations of form (13), all three components of rigid-body rotation found do not differ significantly from zero: $(\omega_x,\omega_y,\omega_z)=(4,7,11)\pm(5,5,5)$~km s$^{-1}$ kpc$^{-1}$. It is possible that for the
application of the graphical method in the analysis of the kinematics of the TW Hya association, even more accurate radial velocities of stars are required, on which the accuracy of the spatial velocities $U, V,W$ mainly depends.

 \subsubsection*{ACKNOWLEDGMENTS}
The author is grateful to the reviewer for useful comments that contributed to improving the work.

 \subsubsection*{FUNDING}
The work was carried out at the expense of the organization's budget.

 \subsubsection*{REFERENCES}
 \small

\quad~1. J. J. Armstrong, J. C. Tan, N. J.Wright, et al., Monthly Notices Royal Astron. Soc. 543 (3), 2349 (2025).

2. A. Blaauw, Annual Rev. Astron. Astrophys. 2, 213 (1964).

3. V. V. Bobylev and A. T. Bajkova, Astronomy Letters 49 (7), 410 (2023).

4. V. V. Bobylev and A. T. Bajkova, Astrophysical Bulletin 79 (3), 473 (2024).

5. R. de la Reza, E. Jilinski, and V. G. Ortega, Astron. J. 131 (5), 2609 (2006).

6. R. de la Reza, C. A. O. Torres, G. Quast, et al., Astrophys. J. 343, L61 (1989).

7. J. K. Donaldson, A. J. Weinberger, J. Gagn\'e, et al., Astrophys. J. 833 (1), article id. 95 (2016).

8. C. Ducourant, R. Teixeira, P. A. B. Galli, et al., Astron. and Astrophys. 563, id. A121 (2014).

9. J. Gagn\'e, O. Roy-Loubier, J. K. Faherty, et al., Astrophys. J. 860 (1), article id. 43 (2018).

10. J. Gregorio-Hetem, J. R. D. Lepine, G. R. Quast, et al., Astron. J. 103, 549 (1992).

11. K. L. Luhman, Astron. J. 165 (6), id. 269 (2023).

12. V. V. Makarov and C. Fabricius, Astron. and Astrophys. 368, 866 (2001).

13. E. E. Mamajek, Astrophys. J. 634 (2), 1385 (2005).

14. A. M. Mel'nik and A. K. Dambis, Monthly Notices Royal Astron. Soc. 472 (4), 3887 (2017).

15. A. M. Mel'nik and A. K. Dambis, Astronomy Reports 62 (12), 998 (2018).

16. N. Miret-Roig, J. Alves, S. Ratzenb\"ock, et al., Astron. and Astrophys. 694, id. A60 (2025).

17. N. Nagananda, L. Vican, B. Zuckerman, et al., Open Journal of Astrophysics 7, id. 80 (2024).

18. K. F. Ogorodnikov, Dynamics of Stellar Systems (Pergamon, Oxford, 1965).

19. J. Olivares, N. Miret-Roig, P. A. B. Galli, and H. Bouy, Astron. and Astrophys. 699, id. A122 (2025).

20. L. Posch, J. Alves, N. Miret-Roig, et al., Astron. and Astrophys. 693, id. A175 (2025).

21. T. Preibisch and H. Zinnecker,Astron. J. 117 (5), 2381 (1999).

22. R. Sch\"onrich, J. Binney, and W. Dehnen, Monthly Notices Royal Astron. Soc. 403 (4), 1829 (2010).

23. A. Vallenari et al. (Gaia Collab.), Astron. and Astrophys. 674, id. A1 (2023).

24. N. J. Wright, New Astronomy Reviews 90, article id. 101549 (2020).

25. B. Zuckerman and E. E. Becklin, Astrophys. J. 406, L25 (1993).

\end{document}